# Towards the Right Direction in BiDirectional User Interfaces

**Yulia Goldenberg**

Software and Information Systems Engineering, Ben-Gurion University of the Negev, Beer Sheva, Israel, yuliago@post.bgu.ac.il

**Noam Tractinsky**

Software and Information Systems Engineering, Ben-Gurion University of the Negev, Beer Sheva, Israel, noamt@bgu.ac.il

Hundreds of millions of speakers of bidirectional (BiDi) languages rely on writing systems that mix the native right-to-left script with left-to-right strings. The global reach of interactive digital technologies requires special attention to these people, whose perception of interfaces is affected by this script mixture. However, empirical research on this topic is scarce. Although leading software vendors provide guidelines for BiDi design, bidirectional interfaces demonstrate inconsistent and incorrect directionality of UI elements, which may cause user confusion and errors.

Through a websites' review, we identified problematic UI items and considered reasons for their existence. In an online survey with 234 BiDi speakers, we observed that in many cases, users' direction preferences were inconsistent with the guidelines. The findings provide potential insights for design rules and empirical evidence for the problem's complexity, suggesting the need for further empirical research and greater attention by the HCI community to the BiDi design problem.

## 1 INTRODUCTION

The global proliferation of interactive technologies and the resulting need to localize products, applications, and websites, has brought to the surface the need to support more than half a billion "Bidirectional (BiDi) language" speakers (e.g., Arabic, Hebrew, Urdu, see [16]). Those languages rely on unique BiDi writing systems, mixing the native right-to-left (RTL) script with left-to-right (LTR) strings - Latin script, numbers, and other notations [6, 12]. An example of BiDi text in Arabic can be seen in Figure 1. In this example, the general text is written in Modern Arabic in RTL direction whereas the name in English, the number, and the years are written from LTR (marked in yellow).

BiDi languages are different not just from LTR languages in their script directionality usage, but also from each other in some rules regarding specific information representation. For example, the appropriate direction for a numerical order (e.g., "123") in Hebrew and Persian (when using Eastern Arabic-Indic numeral system) is ascending LTR, while in Mashreq (Eastern) Arabic the proper direction is RTL [33, 36].

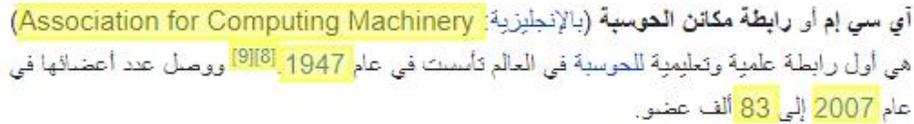

Figure 1. **An example of BiDi text. Text translation: "ACM - The Association for Computing Machinery is the first scientific and educational association for computing in the world, which was founded in 1947. Its membership in 2007 reached 83 thousand members". Retrieved 7 September 2020, from the Arabic version of Wikipedia.org website**

The simultaneous use of two opposite directions over time forms unique perception and behavior habits of "BiDi language" speakers. Among them, for example, is the mirrored Z-shaped reading pattern [20], the right-side saccade preference in the visual field [46], and the "jumping" eye movements, which are associated with the transition from one direction to another within the same sentence, which can be seen in Figure 2.

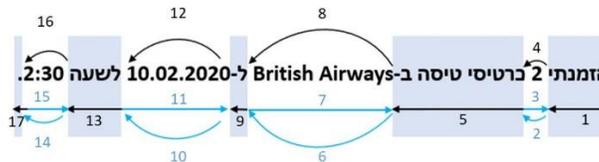

Figure 2. **A schematic depiction of eye-movements while reading the BiDi sentence (on Hebrew) "I booked two tickets at British Airways for the 10th of February 2020 at 2:30". Black lines represent eye movements required for RTL reading; blue lines represent movements required for LTR reading. Top and bottom line locations were used for better readability. Straight lines represent text-reading movements; arched lines represent "instrumental" movements**

Due to the specifics of BiDi languages and the cognitive characteristics of BiDi, the main challenge when localizing or developing a BiDi user interface (UI) is to determine the direction of specific UI elements [52]. Should it appear in a LTR direction, RTL orientation, or should it mix elements with both directionalities in its structure? This problem is especially relevant when companies are creating design systems for BiDi interfaces or for interfaces intended to be localized in BiDi languages. Various organizations have enlisted in response to this challenge. The World Wide Web Consortium (W3C) has provided recommendations for developers and designers working with RTL and BiDi scripts [32, 33, 58], and Unicode Technical Committee developed standards for bidirectional sites [12]. Leading global software companies such as Microsoft [38, 41, 42], Google [24], and Apple [5] created guidelines for designers and developers who are involved in the process of application localization, bidirectional interface creation, and BiDi interface support.

Despite these efforts, incorrect and inconsistent usage of UI elements' directionality abound. For example, Figure 3 shows two contrasting versions of the star rating system used in localized BiDi (Hebrew) sites. Moreover, inconsistent directionality of UI elements in BiDi interfaces can sometimes be found even in products of the same vendor or even worse, within the same product (see Figure 4).

The problem with using incorrect and inconsistent directionality of UI elements in BiDi interfaces is that it hampers automatic data perception, increases the time needed for learning, understanding, and acting on the interface, induces ambiguity and interpretation errors, and leads to user frustration and dissatisfaction with the interface [48].

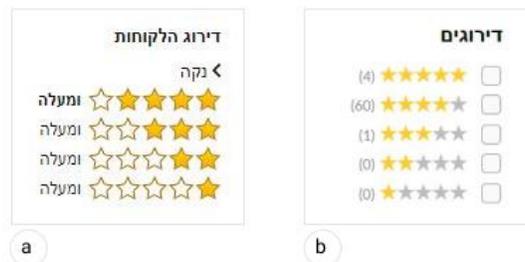

Figure 3. **An example of inconsistent UI element direction usage in BiDi interfaces: (a) RTL star rating from Amazon.com, (b) LTR star rating from iherb.com. Retrieved from Hebrew version web sites on June 9, 2020.**

**Towards the Right Direction in BiDirectional User Interfaces**

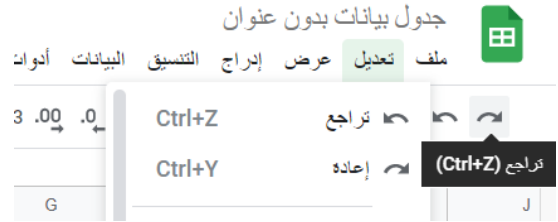

Figure 4. **An example of inconsistent UI element directionality within the same BiDi (Arabic) interface: two functionally identical pairs of icons - "Redo" and "Undo" located in different interface logic blocks (on the left with a label, on the right as an action button with a tooltip), are being used in different directions. Retrieved December 21, 2020, Arabic version of Google spreadsheet web site.**

The reasons for this problem could be traced to misunderstanding the nature of BiDi languages. This is manifested by developers' and designers' tendency to uncritically adopt an "RTL design" approach1 in the interface development process. This approach leads HCI professionals to treat BiDi languages simply as directionally opposite to LTR languages, which entails that all LTR UI elements should be mirrored in the interfaces. Uncritical mirroring provides a fertile ground for incorrect direction usage and is manifested by numerous bugs in BiDi interfaces (see an example of such bug in Figure 5).

Another reason for incorrect UI element direction usage lies in not recognizing the uniqueness of specific BiDi languages described above, which leads to the indiscriminate application of universal BiDi design recommendations to all BiDi languages, regardless of unique (i.e., language-specific) rules for information representation. In most cases breaking these linguistic rules can be easily detected by native BiDi language speakers and guarded against.

Moreover, the directionality of UI elements, such as timelines, charts, and imagery, is not readily governed by linguistic rules, and therefore, it is inappropriate to label usage of such elements as "right" or "wrong." This type of UI elements requires special guidelines to help professionals define the optimal direction of an element and thereby lead users to process information effectively and efficiently. Inconsistent use of those elements, which we observed not only in localized but also in native BiDi interfaces, calls into question the reliability of current BiDi design guidelines, and indicates a need to uncover the reasons for the inconsistent directionality and to find ways to prevent it.

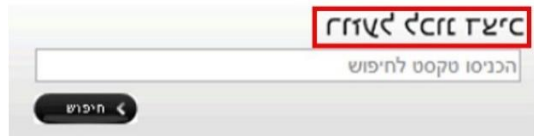

Figure 5. **An example of "blind" mirroring in a BiDi interface. On the figure: instead of using the original direction of the image file with a Hebrew headline (marked in red), the developer flipped it, making a mirror-like letters view and therefore creating an unreadable text. Retrieved on June 9, 2020, from Hebrew version of Next.com web site.**

Thus, we pose the following question: Are current BiDi design guidelines for UI elements appropriate for speakers of BiDi languages? Towards answering this question and improving the justifiability and consistency of BiDi UI design guidelines, this paper provides three main contributions. (1) Reviews of possible reasons for UI element direction inconsistency in top BiDi interfaces (in Hebrew), in current BiDi design guidelines, and in the relevant literature. (2) Mapping of common inconsistent elements in Hebrew BiDi interfaces. (3) An empirical evaluation of BiDi users' and HCI professionals' preferences of UI element directionality and comparison of the results against current BiDi design guidelines. By providing empirical evidence about the BiDi design problem complexity, we aim to raise the HCI community's attention to the BiDi design problem and show the need for further empirical research on this topic.

This paper proceeds as follows: the second section provides a theoretical background; the third section presents the results of an informal review of BiDi (Hebrew) websites and maps a sample of common inconsistent UI elements with potential reasons for their inconsistency. The fourth section describes an exploratory study of UI elements' preferences that demonstrates the problem's complexity. We present the ramifications of our findings in the concluding section.

---

1 The term is frequently used among professional HCI and developers' communities (StackExchange, Smashing Magazine, UXPlanet.org, etc.), as well as in leading tech companies' guidelines [5].

## 2 BACKGROUND

The growth of digital technology, the development of new communication tools, and the proliferation of the internet all over the world in the last three decades have led HCI researchers to study the influence of culture on how users interact with products, applications, and websites (cf., [10, 11, 44]). The challenge of designing interactive technology for BiDi users has been acknowledged since the early 1990s [47]; however, it did not gain popularity in professional circles at that time. The early 2000s are characterized by the lack of BiDi versions for world-known applications [8, 61]. With the increasing number of BiDi users in World Web in the last three decades (for example, the number of Arabic-speaking users has increased by 9,348.0% between 2000 to 2020, which is the highest growth among the top ten web languages [31]), BiDi design has become increasingly popular in professional HCI communities. Leading web companies and individual HCI professionals have been trying to define the optimal direction of UI elements in BiDi interfaces. Both groups rely on the "Partial Mirroring" rule: mirroring just a part of left-to-right UI elements, such as, overall layout, navigation elements, elements with BiDi text, etc., while maintaining the LTR direction for elements that are not related to RTL script, such as numbers, graphs, images without text, media player, etc. [2-5, 21, 24, 41].

Despite extensive discussions on "Bidirectional design" in HCI professional community, there is a paucity of empirical studies on how BiDi users perceive and process the direction of UI elements. Consequently, there is little evidence to rely on for guiding BiDi design and for understanding the basis for recommending the directionality of BiDi UI elements.

In the few empirical HCI studies on BiDi design, researches provide only brief recommendations (e.g. mirror web objects [8, 43]; place important information on the right-side [7]); or define the specifics of BiDi websites (e.g. mirrored layout, objects' right side alignment, RTL navigation orientation, RTL text direction [1, 7, 34, 54]). Most empirical studies related to UI element directionality in BiDi interfaces did not include BiDi users. A notable exception is a study with twenty-five Arabic speaker participants, that has demonstrated a preference for right-hand menus in Arabic websites [55].

However, based on findings from studies in psychology, we argue that empirical evidence for BiDi design guidelines is important. There, the intricacy of BiDi users' perception is demonstrated. These studies show that the general mixed structure of BiDi languages is not only a matter of the directionality of the UI elements because BiDi speakers' direction preferences could be mediated by the influence of various factors, such as linguistic specifics of BiDi language [49, 62], context [18, 53, 56], LTR language proficiency level, number years linguistic specifics of BiDi language of immigration to LTR country or schooling and studying mathematics [13, 49, 53, 62, 64].

These findings portray a picture that challenges the notion of simple, universal transformation of LTR user interface elements to BiDi languages. Furthermore, it emphasizes the need for a systematic study of this domain to augment and improve existing guidelines, rules of thumb, and designers' gut feelings. In the following sections, we aim to achieve preliminary evidence for the potential benefits of such a study.

## 3 REALITY CHECK: HOW PREVALENT IS INCONSISTENT DIRECTIONALITY OF UI ELEMENTS IN BIDI WEBSITES AND WHAT ARE THE POSSIBLE REASONS?

To assess the prevalence and nature of inconsistent UI elements in BiDi websites, we conducted an informal review of the top websites in Israel. The websites were chosen based on Alexa Traffic Rank (September 2019, April 2020) as a commonly used website traffic measurement [59]. We focused on the top 200 websites that had a Hebrew interface, among them: 21% were localized websites (e.g., google.com, youtube.com, facebook.com, et cetera); 79% were local web sites (e.g., edu.gov.il, kan.org.il, ynet.co.il, et cetera). Web sites were represented by the following genres: finance (12%); corporate (8%); e-commerce (21%); education (9%); entertainment (7%); government 7%); news (16%); social media (5%); other genres (15%).

We identified several dozens of UI items that were used in RTL and LTR versions (that is, elements whose directionality was not consistent across the reviewed websites). Of those, we selected 32 UI elements that were used frequently (appeared more than three times in sites of the same genre) for further classification and exploration of the reasons for the inconsistent usage. The review was conducted on Hebrew websites because of our familiarity with this language, but we observed that many similarly "controversial" items could also be found in websites of other BiDi languages. For example, the inconsistent direction usage of the slider [19] or the "Redo" and "Undo" icons (see Figure 4) in Arabic interfaces.

**Towards the Right Direction in BiDirectional User Interfaces**

To help us study the 32 elements, we tentatively classified them into five categories according to the type of their primary perceived information: *quantity, order, and preference* (15 items); *time flow* (1 item); *date* (1 item); *motion* (9 items); *space and navigation* (6 items). Some groups were further divided into sub-groups (see Appendix for a more detailed analysis).

We reviewed potential resources to identify the roots of direction inconsistency for each group, sub-group, or specific item (where possible). These resources included BiDi design guidelines [5, 24, 41, 42] and standards providing recommendations about UI element optimal direction (e.g. [12, 17]); HCI professional literature on BiDi design topic (e.g. [30, 58]); empirical studies on direction preferences of BiDi users published in last three decades (e.g. [18, 62]), and other professional resources.

Comparing the 32 UI elements with the guidelines and the literature on BiDi design, we came to the conclusion that the major reason for the inconsistent directionality of UI elements is the *paucity and incompleteness of current guidelines.* As indicated in the Appendix (fourth column), no specific recommendations were observed for 20 out of the 32 inconsistent UI items.

Moreover, the fact that inconsistency is manifested in elements for which there are design guidelines suggests potential incompatibility of BiDi design recommendations with the direction preference of BiDi users or with the rules of specific BiDi languages. For example, RTL is the recommended direction in BiDi interfaces for a slider (an element, "letting users select a value or range from a fixed set of options" [29]) [26, 40]. However, in Hebrew, LTR is the appropriate direction for numeric values' representation and both LTR and RTL are appropriate directions for writing numeric order and range, which contradicts the guideline.

In turn, the existence of two appropriate directions for the same information representation in the language suggests another possible explanation for the multitude of elements with inconsistent directionality in BiDi interfaces – *that the direction of the element does not really matter.* The results of empirical studies (e.g. [22, 62], see the fourth column in the Appendix for more references) support this point of view. If that is the case, we will likely find the users evenly split in their preference for the RTL or LTR direction of each element. However, we find this explanation unlikely given the analysis of the problem in Section 1(e.g., Fig. 2) and the realization that numerous language-related and contextual factors could affect the BiDi users' direction preferences ([18, 53, 62], et cetera). This issue is further explored in the empirical study (Section 4).

The findings of prevalent inconsistent use of UI element directionality in BiDi websites coupled with the paucity of design guidelines and systematic research on BiDi design indicate a need to gather more evidence regarding this topic. We address this need by studying the preferences of BiDi users regarding the directionality of various UI elements by an online survey with a sample of Hebrew speakers. Based on the findings of our review of the literature (see third and fourth columns of Table 2 in Appendix), we are guided by a general assumption that BiDi users' lack of directional preference of UI elements could explain only some of the identified "inconsistent" UI elements. For many other UI elements, BiDi speakers will have a clear dominant direction preference. More specifically, in the context of Hebrew speakers, we can expect users to exhibit the following directionality preferences.

- An LTR preference for UI elements related to performing mathematical tasks (e.g., counters) and charts. The assumption is based on the fact, that LTR direction usage in math writing in Hebrew.
- An RTL preference for calendars including a Hebrew weekday-weekend format (RTL direction is used for writing Hebrew weekdays) and for icons of objects shown in a side-view profile (e.g., "shopping trolley", "landing aircraft", et cetera) – based on the studies, showing an aesthetic preference for the objects turned to the left [9, 23, 45].
- Our literature review suggests that there is not enough evidence to predict any directionality preference for the other elements.

In addition, we assume that direction preferences can be moderated by contextual factors. For example,

- Users with high proficiency in LTR languages or with professional experience related to interactions with LTR interfaces, documents, or communication with LTR language speakers will prefer LTR element versions, while users with the BiDi language and professional experience will prefer RTL versions (based on the results, showing the influence of LTR language proficiency and the context on direction preferences of BiDi users [49, 53, 62]).
- HCI professionals will be more conscious of the general design guidelines calling for mirroring UI elements and thus will have a stronger preference for RTL direction relative to ordinary users.
- There are underlying logical groups of UI elements that can be discovered from users' directionality preferences (e.g., based on Gestalt principles [63]).

## 4  STUDY OF USER PREFERRED DIRECTIONS

To demonstrate the usefulness of empirical data for the study of designing BiDi UI elements, we developed and administered a questionnaire among two groups of Hebrew-speaking people.

### 4.1  Stimuli

Twenty-seven questionnaire items of image-based UI elements were used in the questionnaire. The elements were based mainly on inconsistent UI elements described in Table 2 (for technical reasons we could not include in the survey all of the identified elements) with the addition of six other UI elements (marked with an asterisk in the Appendix table) that are commonly found in Hebrew interfaces, but which were not observed in our review of the top 200 websites. For the full list of UI items, see Table 2 and "Survey Questions and Stimulus" in supplementary materials.

Five of the items presented a single image of a UI element with potential ambiguous information and asked the participants to choose the correct interpretation of the element (out of two possible interpretations, one based on RTL reading and one on LTR reading). An example is shown in Figure 5a, where the element can be interpreted as "Buy two get three free" if read LTR or "Buy three get two free" if read RTL. In one item we presented as a stimulus, a web page with long text, and asked for the most convenient placement (right side or left side) for the web scroll bar (see an example of used stimuli in the supplementary files). In the other 22 items, we presented an LTR and an RTL version of the same UI element and asked participants which version is more convenient for them while interacting with a website (see an example in Figure 5b).

All UI elements and questionnaire items were formulated in Hebrew and the questionnaires were administered in either mobile or web versions for the users' convenience. File "Survey Questions and Stimulus" in Supplementary materials contains the full list of the questions used in the Survey.

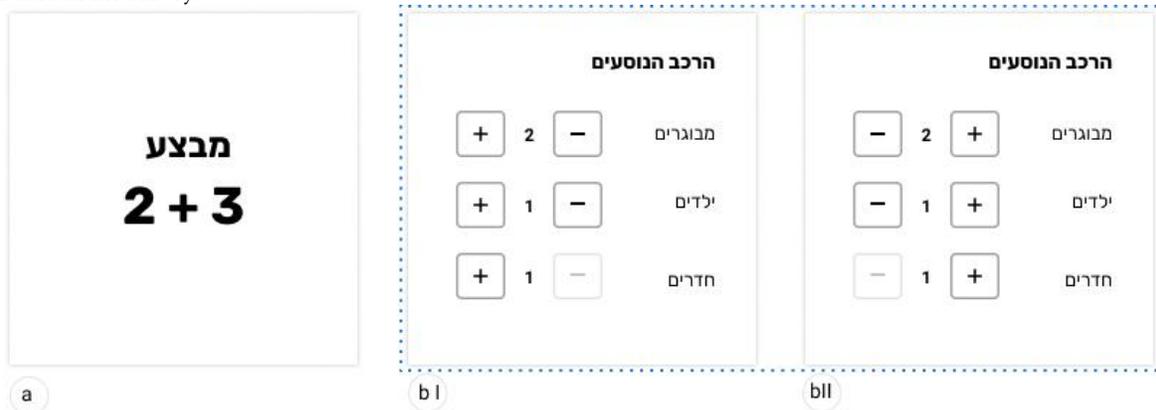

Figure 5. **Examples of the study's stimuli. (a) The UI element with ambiguous information, depending on whether the user reads from right to left or from left to right. The question asks: "From the image, you understand that the offer is...": "Buy two get three free" or "Buy three get two free". (b) A direction preference item (framed in the dashed line). Participants are asked "Which of the two options is more convenient for you to define the number of guests and the number of rooms in the hotel? Click on the more convenient option": (b I) RTL version, (b II) LTR version.**

### 4.2  Procedure

The study was conducted during March and April 2020 using the "Survey Legend" online questionnaire tool. A recruiting post and links for mobile and web versions were published in Hebrew in non-professional groups and HCI professional communities of Facebook and WhatsApp. After clicking on the questionnaire's link, the respondents immediately approached the introduction part and could start their participation. During the study, the participants could see a progress bar (RTL direction, placed vertically at the bottom of the screen). They were not limited by time constraints. Participants were not allowed to fill out the questionnaire more than once. The questionnaires started with a description of the study's goals and the kind of data that will be collected for analysis. This was followed by seven to seventeen (depending on participants' answers) background items. The questionnaire's main part's items were then presented one by one. In an attempt to register the participants' intuitive reactions, we did not provide an opportunity to revise previous answers. The LTR vs

**Towards the Right Direction in BiDirectional User Interfaces**

RTL direction items were presented in random order as pairs of side-by-side elements (in the web version) or one above the other (in the mobile version).

## 4.3 Participants

A total of 290 speaking participants volunteered to participate in the study. Fifty-six participants (19.3%) did not complete the questionnaire and were excluded from the analysis. The remaining 234 participants average age was 35 years (SD = 7.6; range = 20 to 69).

Most of the participants were native Hebrew speakers (228). Almost all participants (99.1%) had a high degree of LTR language proficiency (higher than "excellent in speaking"); 37.6% of participants described themselves as native speakers of at least one LTR language (English, Russian, French, et cetera); 64.9% reported that they are "excellent in writing, reading, and speaking"; and 17.5% - "excellent in reading and speaking" of at least one LTR language. The professional experience of the participants was diverse. For example, 15 participants worked only with BiDi (Hebrew, Arabic, et cetera) speakers, 23 combined work with BiDi speakers and work with BiDi interfaces; 70 combined work with LTR speakers and work with LTR interface.

There were 105 respondents without HCI professional background, 129 had some HCI-related professional experience (26 had experience mainly in interface programming, 83 in UX design, and 20 in UI design). HCI professionals were engaged in product development in various languages (see Supplementary Materials). HCI professionals were selected as representatives of the HCI community's "common wisdom", who should be familiar with the guidelines and might differ in UI elements' direction preferences from naive users.

One-hundred and eighty-nine participants used the survey's mobile version, while 45 used the web version. We associate the relatively large number of mobile respondents with the study's recruitment channels (Facebook and WhatsApp groups).

## 4.4 Results

The data was analyzed using Python Statsmodels [57] package. A series of Mann-Whitney tests with Bonferroni correction (confidence level 99.8%) tested whether the survey version (web or mobile) affected the UI direction preferences of the participants. No significant difference was found (p > 0.1 for all elements). Consequently, responses from both device types were combined for the ensuing analyses.

### 4.4.1 Preference of UI Elements Direction.

To test for the directionality of preferences of the selected UI elements among all participants, we performed a Z-test for a single proportion (50% is a target value), confidence level 99.8% - calculated with Bonferroni correction. We found that directionality preference differed depending on the UI elements. Out of the 27 elements, 13 were preferred in their LTR version, eight elements were preferred in their RTL version, whereas users did not show clear preference regarding the directionality of only six elements. The detailed data about the directionality preference of the UI elements are shown in Table 1 and Figure 6.

Table 1. The proportion of participants' preference of UI elements' direction (all significant differences are at p<.01 level)

| UI Element | Significant Direction Preference | | | Preferred interpretation (a) |
|---|---|---|---|---|
| | RTL | LTR | No Preference (b) | |
| 1. Star Rating | 62%, z=3.73 | | | Far left is a maximum value, far right is the minimum |
| 2. 1+2 מבצע (c) | 79%, z=8.69 | | | "buy 2, get 1 free" |
| 3. 2+3 מבצע | | 82%, z=9.87 | | "buy 2, get 3 free" |
| 4. 3+1 מבצע | | 85%, z=10.79 | | "buy 3, get 1 free" |
| 5. Time range | | 80%, z=9.22 | | "Parking is free from 09:00 till 21:00" |
| 6. Numeric Rating | | 68%, z=5.43 | | |
| 7. Counter | | 65%, z=4.51 | | |
| 8. Slider | | 90%, z=12.09 | | |
| 9. Price range | | 70%, z=5.95 | | |
| 10. Size range | | 84%, z=10.39 | | |
| 11. Year range | | 87%, z=11.18 | | |

| UI Element | Significant Direction Preference | | | Preferred interpretation (a) |
|---|---|---|---|---|
| | RTL | LTR | No Preference (b) | |
| 12. Bar chart | | 68%, z=5.30 | | |
| 13. Trend Graph | | 91%, z=12.49 | | |
| 14. Timeline | | 69%, z=5.82 | | |
| 15. "Valid card Date" input | | | LTR=58%, z=2.42 | |
| 16. Calendar with Hebrew weekdays | | | RTL=54%, z=1.24 | |
| 17. Calendar without weekdays | | | LTR=57%, z=2.16 | |
| 18. Calendar with BiDi weekdays | | | LTR=59%, z=2.55 | |
| 19. Accessibility icon | 80%, z=9.09 | | | |
| 20. Arrival icon | 70%, z=5.95 | | | |
| 21. Bus icon | 82%, z=10.00 | | | |
| 22. Departure icon | | | RTL=55%, z=1.50 | |
| 23. Phone icon | 71%, z=6.34 | | | |
| 24. Send icon | 65%, z=4.51 | | | |
| 25. Shopping trolley icon | 84%, z=10.39 | | | |
| 26. Inside scroll bar | | | RTL=54%, z=1.11 | |
| 27. Web scroll bar | | 69%, z=5.69 | | |

Notes: (a) For UI elements including possible ambiguous information; (b) Data about the largest proportion is shown in the table; (c) For items 2-4, the strings are of the type "Sale x+y".

### 4.4.2 Influence of personal factors.

We performed a logistic regression analysis with Maximum likelihood estimation (MLE) to evaluate the influence of LTR language proficiency and of working characteristics (with LTR or BiDi interface, documents, or speakers) on the respondents' UI element direction preferences. We did not include in the analysis RTL language proficiency covariate since a major part of our participants (97.4%) were native Hebrew speakers. In our model, we used 7 predictors: proficiency in LTR languages (maximal proficiency level of any of the LTR languages); professional experience related to interaction with LTR or BiDi interfaces, documents, or communication with LTR or BiDi language speakers, and a bias term. The dependent variable was the direction preference of the UI element of each of the answers. The model was not significant ($p>0.05$), and none of the covariates' coefficients was significant ($p>0.05$). The bias term was -0.412 ($p=0.02$), due to an overall bias to LTR answers for the selected set of elements.

### 4.4.3 Influence of HCI professional experience.

To test whether there are differences in the preferences of ordinary BiDi users and BiDi HCI professionals, we performed a set of Mann-Whitney tests (existence of HCI professional background as an independent variable, UI elements' direction preference as the dependent variable, confidence level 99.8% - calculated with Bonferroni correction). Only one element, the "Bus icon" exhibited such a difference (U = 5713.5, $p < 0.001$). Participants with HCI professional experience demonstrated a higher RTL direction preference for the bus icon (90% vs 74%). However, no such differences were found for the other elements (for more data, see **Figure 6**).

However, the overall picture is more subtle. While not statistically significant in most cases, HCI professionals seemed to have an overall lesser tendency to prefer the LTR direction for UI elements (55%) relative to ordinary users (59%). By comparing the average preference of participants across all UI elements, a Mann-Whitney test (existence of HCI professional background as an independent variable, average UI elements' direction preference as a dependent variable) revealed a significant difference (U= 4733316.0, $p<0.001$) between these two categories of respondents.

**Towards the Right Direction in BiDirectional User Interfaces**

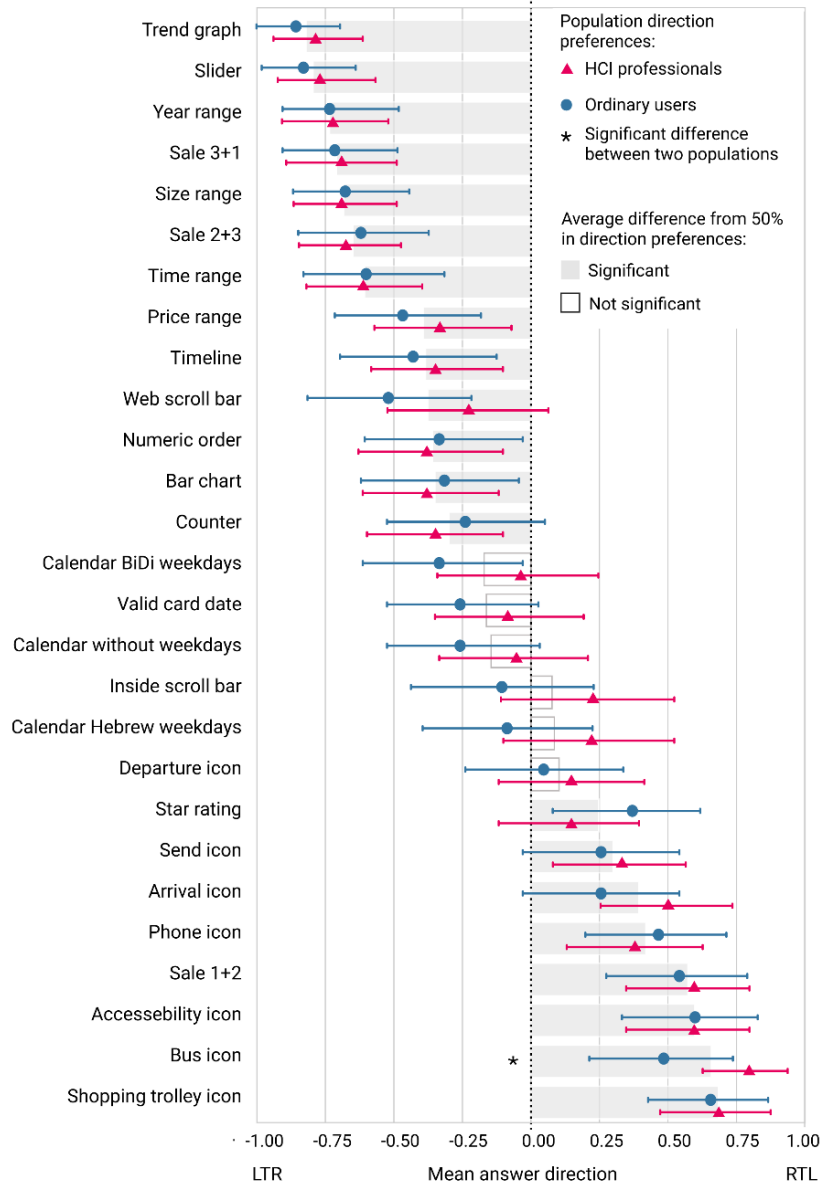

Figure 6. **Average UI elements' direction preferences of HCI professionals (in red with triangular) and ordinary users (in blue with circle), and general average (background bars). Values from -1 to 0 represent LTR direction preference, values from 0 to 1 - RTL direction preference. Confidence level calculated with Bonferroni correction (99.8%).**

**4.4.4 Clusters of UI element information type.**

Finally, we explored the existence of UI elements groups based on the participants' preference for those elements' directionality. With scikit-learn machine learning library for Python [50], we ran a clustering algorithm to group answers according to the patterns of the participants' preferences. The algorithm, unsupervised clustering method (K-Means), was based on the distance between the components, according to participants' answers. Elements with high similarity in responses were assigned to the same cluster. The number of the clusters was chosen by using the Elbow Method [35] (K=5, SSW=322108.3, SSB =3788050.6).

To visually analyze the similarity and variance between the components, we ran a T-distributed Stochastic Neighbor Embedding (t-SNE) [37] technique and reduced the dimensionality from 234 participants into a two-dimensional space which is plotted in Figure 7. The results support the notion that UI elements were grouped according to their directionality preference. Moreover, four out of the five information type groups described in our preliminary review (Section 3) were identified as-is in this clustering analysis.

- The blue circles' cluster in Figure 7 corresponds to the sub-group "Icons of objects, shown in profile" from the "Motion" group.
- The orange squares cluster corresponds to elements from the "Quantity, Order and Preference" group, excluding the star rating and the "Sale 1+2" elements.
- The green triangles cluster corresponds to the sub-group "Calendars" ("Date" group).
- The red diamonds cluster corresponds to elements from the sub-group "Vertical scroll bars" from the "Space and navigation" group.

Three of the 27 elements did not cluster according to our initial classification. The fifth cluster, (purple stars), includes two elements from the group "Quantity, Order and Preference." These are the star rating (sub-group "Numerical order") and "Sale 1+2" (sub-group "Containing the offer "Sale "X+Y"). In addition, the timeline element from the "Time flow" group ("Timeline" sub-group) was associated with the orange squares cluster and demonstrated similar directionality preferences as elements of the "Quantity, Order and Preference" group.

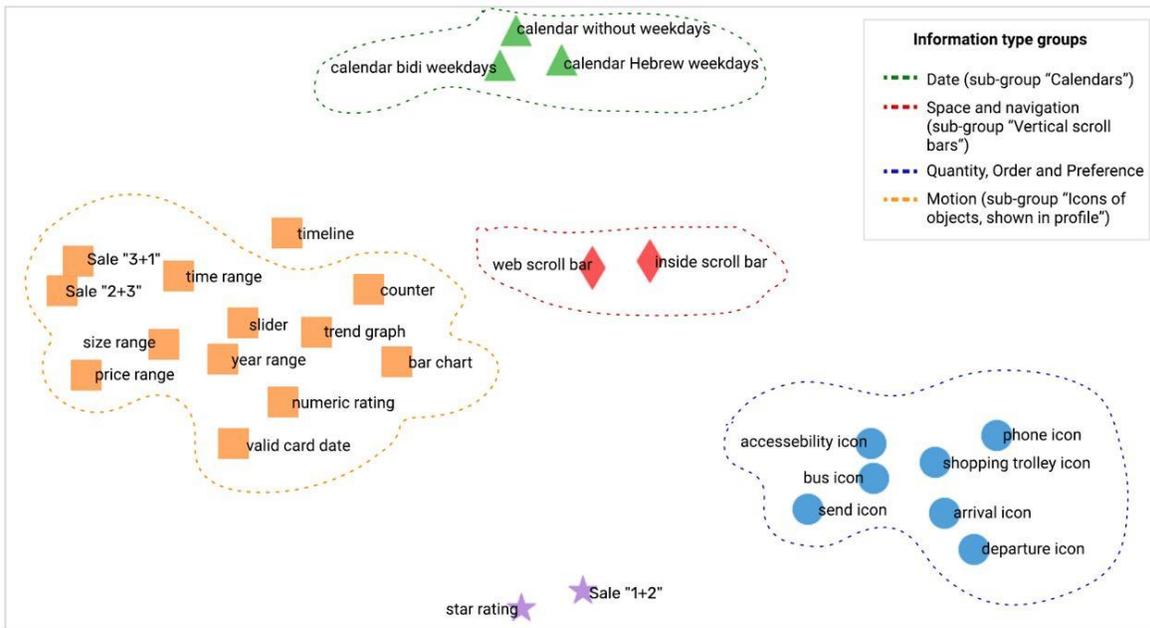

Figure 7. **Two-dimensional t-SNE visualization of UI elements and K-Means clustering (K=5)**

## 4.5 Discussion, Limitations, and Future Research

Common advice about BiDi design is that "UIs for languages that are read from right-to-left (RTL), such as Arabic and Hebrew, should be mirrored to ensure content is easy to understand" [24]. At least regarding UI elements that are not purely textual content, the study's findings suggest a more nuanced view of BiDi design, one that is more complex and multi-faceted. This study demonstrates that Hebrew speakers have differential preferences for the directionality of various UI elements. Moreover, the direction preference of Hebrew speakers corresponded to the recommended direction only for 3 UI elements out of 7 mentioned in the guidelines (for more information see the last column in the Appendix). Very few of UI elements approached consensus about their directionality in a Hebrew interface. Across all selected items, our respondents tended to prefer the LTR version of the UI elements over the supposedly more appropriate RTL version. Although the latter finding may be an artifact of the mix of elements used in our study, we consider these findings as evidence for the need to study BiDi design more systematically and to gather empirical evidence upon which to base future design guidelines.

Because this was an exploratory study, we do not have enough information to explain some of the findings regarding specific elements. Thus, many such findings require further investigation. For example, the lack of direction preference for all types of calendar elements raises a question about how users perceive the nature of this element and its subcomponents. Do users experience an increased cognitive load given the combination of numerical values (dates), which are naturally written from LTR, and the weekday string values, which are naturally written from RTL?

**Towards the Right Direction in BiDirectional User Interfaces**

An interesting finding relates to the fact that the participants preferred the browser's scroll bar in its LTR version (i.e., the scroll bar is on the right-hand side of the window, as is customary in LTR languages) but were split in their preferences regarding the inside scroll bar (i.e., scroll bars within the web site). This can be explained by the abundant experience of Hebrew speakers with the right-side location of the browser's vertical scroll bar (from browsing LTR websites and from using Hebrew websites that maintain the right-side location of the scroll bar). However, the inner element may be perceived differently. For example, its use may be done locally (for a specific element rather than for the entire site), or users' preferences regarding the inner scroll bar may have been triggered more by reflection than by habit. Again, such questions should be investigated in future studies.

Another puzzling finding is the lack of direction preference for the "departure" icon in conjunction with a significant RTL direction preference for the "arrival" icon (see Figure 8). This finding could have several explanations. These two elements could be perceived not just as icons of the same logic group – "spatial mapping" icons [30], nor just as the same object – "airplane" -- differing only in the angle of cockpit rotation, but also as different items that relate to the participants' experience of seeing them as pictograms in international airports. In this vein, a participant wrote to us: "*On the one hand, I prefer to see the icons' direction according to the text direction (e.g., airplane takes off to the left), but on the other hand, I am used to seeing an airplane 'taking off to the right' in every airport that I visit, so taking off to the left feels inappropriate.*"

More focused research can examine these explanations (and possible others) more thoroughly, perhaps in conjunction with the study of other images that point up and down. Similarly, the finding of a significant opposite direction for the elements that we considered as belonging to the same logic groups (star and numeric ratings; promotion offers "Sale 1+2" and "Sale 2+3", "Sale 3+1") cannot be resolved without comprehensive investigation of how users perceive and process these elements.

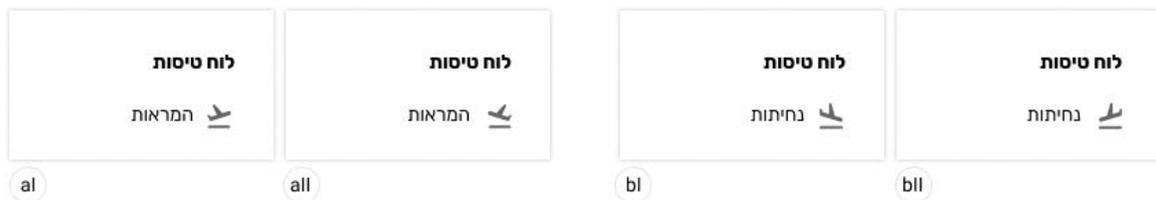

Figure 8. **An example of stimuli used for icons of the same category demonstrated different direction preferences. (a) "Departure" icon (aI) LTR version, (aII) RTL version. (b) "Arrival", (bI) LTR version, (bII) RTL version. The headline in the stimulus reads "Flight schedule", text in the labels: (a) "departures" and (b) "arrivals".**

**4.5.1 UI elements with potential ambiguous information.**

Five of the questionnaire's items included stimuli that had a single image with potentially ambiguous information [items 1-5 in Table 1]. The results indicate that these elements are indeed ambiguous, and as such their usage in Hebrew interfaces in the same form as they were presented in the study, is problematic. For example, the most consensual item of that group was the item labeled "Sale 3+1", which according to the rules of the Hebrew language should have been read "Sale 1+3". Eighty-five percent of the respondents preferred the 3+1 interpretation, probably based on common sense and experience with similar events. However, 15% of the respondents still had the opposite interpretation. This proportion of dissenting interpretations is larger for other elements such as "Sale 2+3" or the star rating. It is therefore important to establish a clear standard for those elements, both to improve the accuracy and to reduce the cognitive effort of the users. Such standards can be based, for example on eye-movement and reaction time data.

**4.5.2 Background factors.**

Interestingly, Hebrew-speaking HCI professionals tended to prefer a more RTL orientation of UI elements relative to ordinary Hebrew-speaking users. This finding may be the result of HCI professionals being more aware of guidelines regarding mirroring UI elements for BiDi sites and applications. However, it also raises the question of whether relying on those guidelines may not eventually increase the gap between the users' mental model of how BiDi interfaces should look like and how such interfaces are designed.

The lack of influence of language proficiency or professional background variables does not necessarily exclude the possibility of their influence. Because the sample in this study was not highly heterogeneous, the effects of those variables may have been limited. This limitation should be addressed in future studies including considering a more refined investigation into aspects such as working background, conventions of online cultural practice that accentuate LTR directionality, and language proficiency. Future research should also look into other background and contextual factors that were not included in this study to form a broader framework to guide research on BiDi UI design.

### 4.5.3 Influence of UI element information type on direction preferences.

The emergence of clusters of UI elements, that are highly similar to our initial classification of element groups, indicates that this classification can serve as a reasonable basis for future research and BiDi design guidelines. This finding suggests that the classification of UI elements by information type is both feasible and potentially useful for devising design guidelines for Hebrew BiDi interfaces. It is also important to validate the preliminary classification presented in this work and to look for other possible element groups that had not been identified yet. Next, it is important to examine further how the information embedded in elements that belong to different classes is processed by BiDi users. Eye-tracking methods appear most suitable for this task. Such studies should integrate potential moderating factors such as the language directionality of the main interface script, the user's task, the user's proficiency in LTR languages, or background in mathematics.

## 5 CONCLUSIONS

There are more than 500 million people in the world who use the BiDirectional writing system. Half a billion users are struggling daily with inconsistent interfaces; Interfaces with no clear standards, where each specialist relies on their own experience and intuition or on guidelines based on insufficient evidence. This work was motivated by our recognition that this state of affairs can be changed for the better. Towards this goal, we contribute to the design of BiDi user interfaces in three aspects. First, we challenged on theoretical grounds the conventional wisdom that the default design of UI elements for a BiDi language should mirror the LTR design of those elements. Second, we demonstrated that in practice there are considerable deviations from BiDi design guidelines and that the directionality of UI elements in major Hebrew websites is inconsistent from site to site and even within sites. Third, findings of a survey conducted with Hebrew speakers augment the first and the second contributions by showing that BiDi professionals and users exhibit a wide range of directional preferences for various UI elements.

We believe that the major source of the inconsistent use of the directionality of UI elements is insufficient theoretical and empirical work. Consequently, there are many UI elements for which there are no guidelines for BiDi design, and existing guidelines may be inadequate for the task. We presented more than two dozen examples of elements that are challenging from a BiDi design perspective and then offered a preliminary classification of those elements, which was largely supported by a cluster analysis of the respondents' preferences.

The psychological literature suggests that various other contextual factors influence users' preferences, thus questioning the universality of BiDi design guidelines. In this study, we found that HCI professionals tend to favor mirroring UI elements more than ordinary users, but we did not find other background or contextual effects. Still, these findings and potential factors that were not examined here should all withstand additional empirical scrutiny.

Another important question is whether these findings, which are based on studying Hebrew speakers, can be generalized to other BiDi languages. Given the overall similarities between BiDi languages (e.g., the similar UI issues in Arabic and Hebrew illustrated in Figures 1 - 4), we believe that the study's major implications apply to most BiDi languages. Yet, this belief should be tested, as there are idiosyncratic language rules that may require fine-tuning some language-specific BiDi guidelines, especially regarding UI elements, including numeric strings. We hope that this work will motivate studies in other languages that will contribute to understanding the problem and to developing design guidelines.

Meanwhile, we recommend that designers pay more attention to the intricacies of BiDi user interfaces and their users, incorporate the issue of UI element directionality in their usability studies, ask for users' feedback, and test their assumptions about element direction. Most importantly, we call designers, usability professionals, and researchers to share their findings and insights with the professional and scientific communities as a means towards broader evidence-based guidelines for BiDi design.

### ACKNOWLEDGMENTS

We thank Dima Goldenberg for his invaluable help in data processing, Mohsen Abdalla and Razy Alchech for their help in preparing examples in Arabic, and the anonymous reviewers for valuable feedback.

**Towards the Right Direction in BiDirectional User Interfaces**

**Towards the Right Direction in BiDirectional User Interfaces**

## A. APPENDICES

### A.1. Identified UI items with inconsistent direction usage in Hebrew interfaces, directionality guidelines, findings from the literature, common usage in Hebrew, and study findings.

**Table 2:** Identified UI item with inconsistent direction usage in Hebrew interfaces and their preferences

| Identified UI item | Recommended direction in the guidelines | Direction preference of BiDi users observed in psychological empirical studies | Acceptable or commonly used direction in the Hebrew language | Current study preference |
|---|---|---|---|---|
| **Group "Quantity, Order and Preference"** | | | | |
| *Sub-Group "Numerical Order"* | | | | |
| Star Rating | -- | BiDi participants preferred both RTL numerical order (ascending) [28, 56, 64] and LTR order [53, 65]. Hebrew speakers in the study of quantity (more-less) and preference (worse-better) demonstrated LTR, bottom-to-top, and RTL (slightly) direction preferences [62]. | Both LTR and RTL options for writing numeric order are used. LTR - according to mathematical order, RTL - according to Hebrew writing numbers in the text of type "1,2,3 and 4". | RTL |
| Paginator | -- | | | N/A |
| Numeric Rating[*2] | -- | | | LTR |
| *Sub-Group "Performing mathematical tasks"* | | | | |
| Counter | -- | -- | LTR writing direction is used in mathematical tasks and for numerical values. | LTR |
| Slider | RTL [26, 40] | -- | | LTR |
| *Sub-Group "Containing the offer "Sale "X+Y"* | | | | |
| מבצע 1+2 | -- | -- | Acceptable both options: LTR - according to mathematical notation addition, RTL - according to the main text direction in Hebrew. | RTL |
| מבצע 2+3 * | -- | -- | | LTR |
| מבצע 1+3 * | -- | -- | | LTR |
| *Sub-Group "Containing range string"* | | | | |
| Date range | -- | -- | There is a lack of consensus about writing a range string in Hebrew. RTL - according to the decision of the Academy of the Hebrew Language [60]. LTR – according to a major group of Hebrew speakers [51]. | N/A |
| Minute range | -- | -- | | N/A |
| Size range | -- | -- | | LTR |
| Temperature range | -- | -- | | N/A |
| Time range | -- | -- | | LTR |
| Year range | -- | -- | | LTR |
| Price range * | -- | -- | | LTR |
| *Sub-Group "Graphs and charts"* | | | | |
| Bar chart | LTR [5, 24] | Potential references are the same, described for numerical order sub-group. | LTR axes' view (y-axis is on the left, x-axis runs from LTR) is used in math writing in Israel. | LTR |
| Trend Graph | LTR [5, 24] | | | LTR |
| *Sub-Group "Hierarchical order"* | | | | |
| "Valid card Date" input | -- | -- | Acceptable both options: LTR (month on the left, year on the right) – according to the credit card view; RTL - according to the Hebrew text typing. | NS |

---

[2] Here and below – an item, marked with the asterisk was not observed as "problematic" during website review, and was included to the exploratory study as commonly used (appears more than three times in sites of the same genre) in the Hebrew websites.

| Identified UI item | Recommended direction in the guidelines | Direction preference of BiDi users observed in psychological empirical studies | Acceptable or commonly used direction in the Hebrew language | Current study preference |
|---|---|---|---|---|
| **Group "Time flow"** | | | | |
| Timeline | RTL [5, 24] | Preference both for LTR and RTL temporal order was found among Hebrew speakers [20, 62] | -- | LTR |
| **Group "Date"** | | | | |
| Calendar with Hebrew weekdays | RTL [17] | -- | RTL | NS |
| Calendar without weekdays * | -- | -- | LTR | NS |
| Calendar with BiDi weekdays * | -- | -- | LTR or RTL | NS |
| **Group "Motion"** | | | | |
| *Sub-group "Action Icons"* | | | | |
| Backward | RTL [24, 27, 30] | RTL (aesthetic preference for the objects turned to the left) [9, 23, 45]. | -- | N/A |
| Redo | RTL [24, 27, 30] | | -- | N/A |
| Undo | RTL [24, 27, 30] | | -- | N/A |
| *Sub-group "Icons of objects, shown in profile"* | | | | |
| Accessibility | -- | RTL (aesthetic preference for the objects turned to the left) [9, 23, 45]. | -- | RTL |
| Arrival | RTL [27] | | -- | RTL |
| Bus | -- | | -- | RTL |
| Delivery | -- | | -- | N/A |
| Departure | RTL [27] | | -- | NS |
| Phone | -- | | -- | RTL |
| Shopping trolley | -- | | -- | RTL |
| Send * | -- | | -- | RTL |
| **Group "Space and Navigation"** | | | | |
| *Sub-group "Buttons"* | | | | |
| "Back", "Next" buttons order | RTL [24] | -- | -- | N/A |
| Close button in dialog | -- | -- | -- | N/A |
| Primary and secondary buttons' order | RTL [24][3] | -- | -- | N/A |
| Switcher (toggle) | -- | -- | -- | N/A |
| *Sub-group "Vertical Scrollbars"* | | | | |
| Browser scrollbar | -- | Not empirical studies: right-side position (as in LTR interfaces) providing support for right-handed users [14]. The left-side position preserves the visual clearness in the primary side of interface content [15]. | -- | RTL |
| Inside-element scrollbar | -- | | -- | LTR |

---

[3] It should be noted that there is a general inconsistency in primary and secondary buttons' order definition (independent of the BiDi interface). For example, Google recommended to show "cancel" button as a first action button in the dialog [25], while Microsoft uses it as a second [39].